\documentclass[final,1p,times]{elsarticle} 
\usepackage{graphicx} 
\usepackage{amssymb} 
\usepackage{amsthm} 
\usepackage{lineno} 

\journal{Nuclear Physics A} 
\begin{document} 

\begin{frontmatter} 


\title{ Measurements of Fragmentation Photons with the PHENIX Detector}

\author{Ali Hanks$^{a}$ for the PHENIX Collaboration}

\address[a]{Columbia University, Nevis Labs, 
P.O. Box 137
Irvington, NY, 10533, USA}

\begin{abstract} 
Direct photons associated with jets provide a direct measurement of the effects of energy loss on the fragmentation of the parton as it propagates through the medium. Perturbative QCD calculations describe the direct photon cross section well at next-to-leading order, predicting a significant contribution from photons produced through parton fragmentation. Non-perturbative quantities such as the photon fragmentation function, which is poorly constrained, lead to large theoretical uncertainties. The measurement of photons correlated with jets in $\mathrm{p} + \mathrm{p}$ collisions serves as an important test of these calculations and is an essential baseline measurement for comparison to $\mathrm{A} + \mathrm{A}$ collisions. A natural way of selecting such photons is to study hadron-photon correlations. Results for the production of photons associated with high $p_{T}$ hadron triggers are presented for PHENIX $\mathrm{p} + \mathrm{p}$ data at 200 GeV center-of-mass energy. 
\end{abstract} 

\end{frontmatter} 



The energy loss of jets as they propagate through the medium produced in heavy ion collisions is thought to be largely due to medium induced gluon bremsstrahlung \cite{white}. However, because the radiated gluons interact with the medium, the radiation spectrum cannot be measured directly. Models attempting to describe the mechanisms for jet energy loss rely on a variety of assumptions including the thickness of the medium and the energy of the parton. It would be useful to have a direct probe of the radiating parton at all stages in its evolution. In addition to gluons, photons are produced through jet fragmentation, as well as induced bremsstrahlung emission in the presence of a medium \cite{enhance}. However, once produced they do not interact strongly with the medium, and would therefore provide just such a probe.

In p+p collisions, NLO perturbative QDC calculations for the direct photon spectra agree well with the data \cite{photons}. At NLO, these pQCD calculations include a contribution from photons produced through parton fragmentation of as much as $20-30\%$ at low $p_{T}$ \cite{vitev}\cite{incnlo}. However, previous measurements of the photon fragmentation function using e+/e- data have been unable to constrain calculations \cite{fragfunc}, leading to large theoretical uncertainties. Comparing the direct photon cross section with and without an isolation cut, thereby removing much of the fragmentation photon component, has also provided little constraint for theoretical calculations \cite{photons}. Measuring fragmentation photons directly therefore provides powerful constraints of pQCD predictions, as well as a baseline for similar $A+A$ measurements.

Because fragmentation photons will be strongly correlated with the high-$p_{T}$ parton that produced them, a natural way of selecting directly for fragmentation photons is to measure hadron-photon correlations and extract the yield for associated photons on the near side. This naturally exludes other sources of direct photons, which are not produced as part of the jet, and should therefore have no correlation with the hadron. This method can then be extended to obtain additional information about the distribution of photons within the jet, such as $j_{T}$, to further constrain theoretical descriptions of parton fragmentation.

\begin{figure}[ht]
\centering
\includegraphics[scale=0.35]{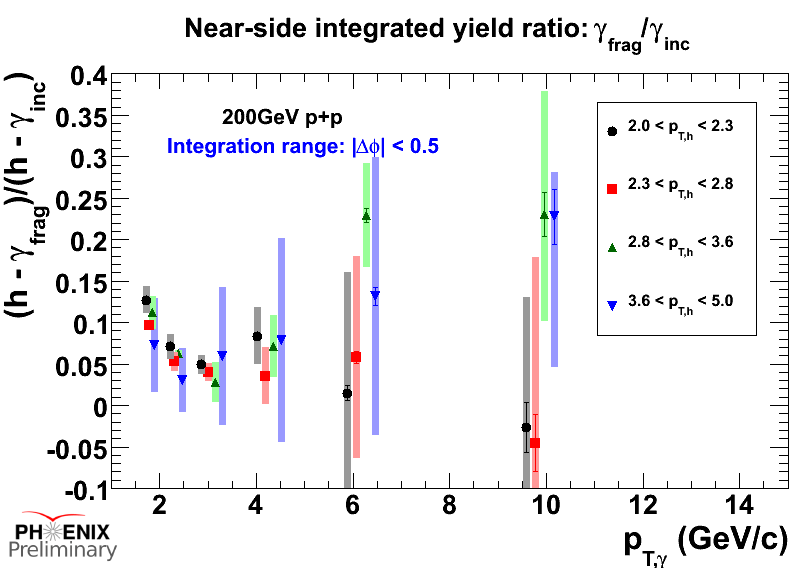}
\caption[]{Ratio of the integrated per trigger yield for fragmentation photons to that of all photons, in a fixed $\Delta\phi$ range.}
\label{yield_ratio}
\end{figure}

The inclusive hadron-gamma per trigger yield (see Eq. \ref{eq:inc}) will be dominated by decay photons, produced through $\pi^{0}$, $\eta$ and heavier mesonic decays. The full decay background is determined by tagging photons that resulted from the decay of a $\pi^{0}$ or $\eta$ directly (see Eq. \ref{eq:dec}). The efficiency, $\varepsilon(\Delta\phi)$, with which such photons are tagged is then determined using simulations that compare the yield of photons correctly tagged as coming from a decay to the yield of all true decay photons, as a function of their position in the detector. The contributions from $\eta$ decays is used to estimate the remaining background using $R_{h/\eta}$, the ratio of $\eta$ plus heavier hadronic decay photons to $\eta$ decay photons, as determined from simulation.

\begin{equation}\label{eq:inc}
\frac{1}{N^{h}_{trig}}\frac{dN^{h-\gamma_{inc}}}{d\Delta\phi} = 
\frac{1}{N^{h}_{trig}}\frac{dN^{h-\gamma_{frag}}}{d\Delta\phi}+
\frac{1}{N^{h}_{trig}}\frac{dN^{h-\gamma_{dec}}}{d\Delta\phi}
\end{equation}

\begin{equation}\label{eq:dec}
\frac{1}{N^{h}_{trig}}\frac{dN^{h-\gamma_{dec}}}{d\Delta\phi} =  
\frac{1}{\varepsilon(\Delta\phi)}
\frac{1}{N^{h}_{trig}}\frac{dN^{h-\gamma_{\pi^{0}-tag}}}{d\Delta\phi} + 
R_{h/\eta}\frac{1}{\varepsilon(\Delta\phi)}
\frac{1}{N^{h}_{trig}}\frac{dN^{h-\gamma_{\eta-tag}}}{d\Delta\phi}
\end{equation}

From this, the integrated per trigger yields for photons within a fixed angular range of the trigger hadron can be determined, and compared to that for inclusive photons, to quantify the strength of the fragmentation photon signal. Figure \ref{yield_ratio} shows the resulting fractional yield, for several ranges of hadron trigger $p_{T}$. The result shows a significant yield of fragmentation photons at low $p_{T}$, with systematic errors due to background dominating at high $p_{T}$. Although not directly comparable to theoretical predictions involving invariant cross-sections, rather than conditional yields, these results are consistent with what is expected based on such calculations \cite{vitev}\cite{incnlo}, within the current systematic limitations.

\begin{figure}[ht]
\centering
\includegraphics[scale=0.35]{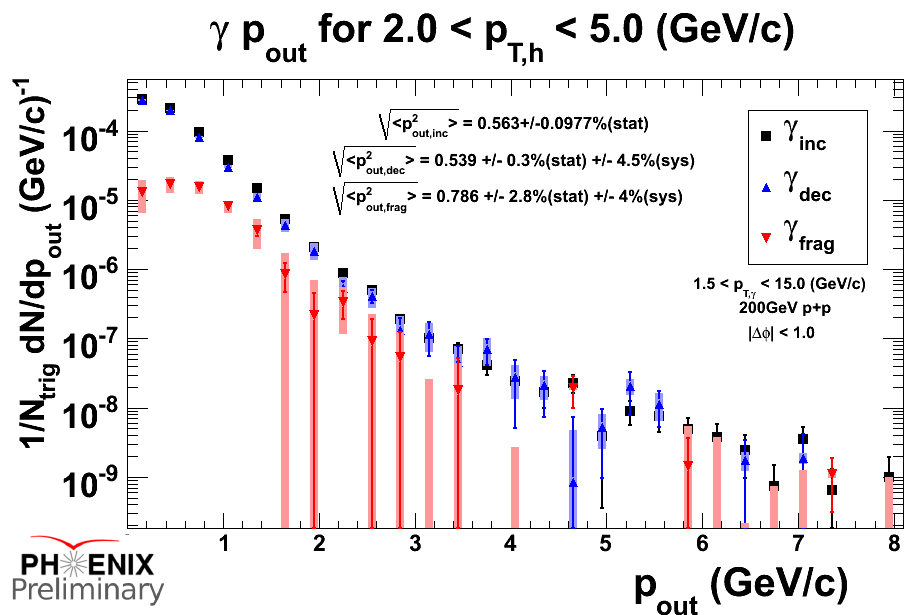}
\caption[]{The $p_{out}$ distribution for inclusive, decay, and fragmentation photons associated with trigger hadrons.}
\label{pout_total}
\end{figure}

\begin{figure}[ht]
\centering
\includegraphics[scale=0.45]{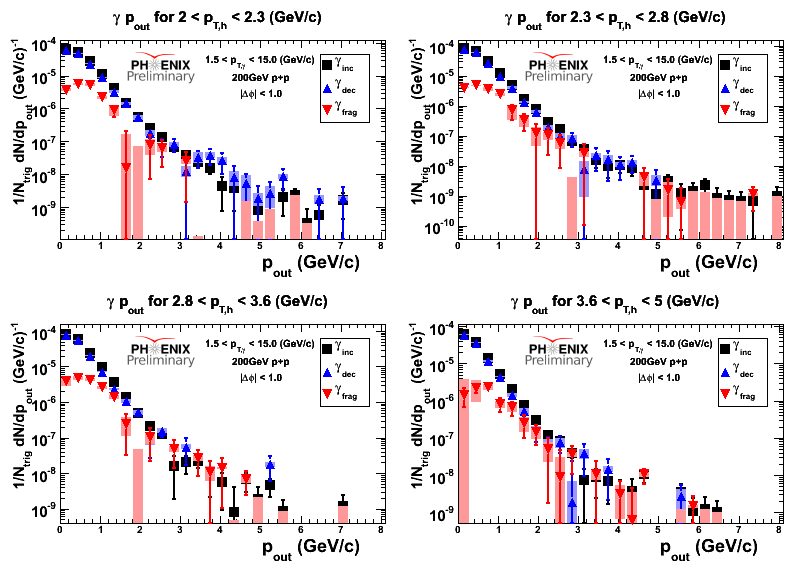}
\caption[]{The $p_{out}$ distributions for inclusive, decay, and fragmentation photons associated with trigger hadrons with $p_{T,t}$in several trigger $p_{T}$ bins.}
\label{pout_bins}
\end{figure}

Further details of jet fragmentation can be obtained by studying the jet properties of these fragmentation photons, such as $p_{out}$, the component of the photon momentum perpendicular to the trigger hadron. When looking at photons in the same jet as the trigger hadron, this is equivalent to $j_{T}$. Figures \ref{pout_total} and \ref{pout_bins} show the resulting $p_{out}$ distributions for inclusive photons (squares), decay photons (up triangle), and fragmentation photons (down triangle). The photon $p_{T}$ range from $1.5 - 15.0~\mathrm{GeV/c}$, while the hadrons are broken into four bins ranging from $2$ to $5~\mathrm{GeV/c}$, with Figure \ref{pout_total} showing the four bins combined. 

\begin{figure}[ht]
\centering
\includegraphics[scale=0.35]{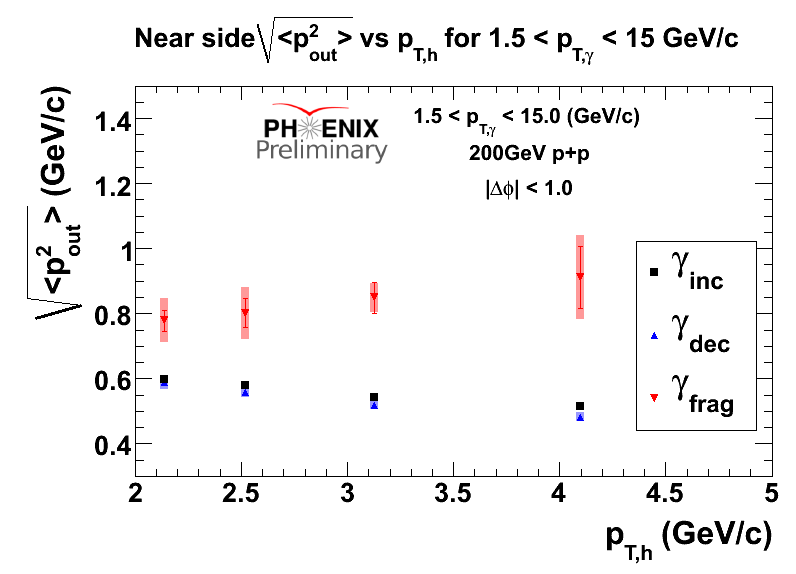}
\caption[]{The $p_{out}$ rms values vs trigger hadrons $p_{T}$ for inclusive, decay, and fragmentation photons.}
\label{pout_rms}
\end{figure}

We see a clear qualitative difference in the shape of these distributions for fragmentation photons, as compared to the inclusive or decay photon distributions. We can quantify this by measuring the rms, defined as the standard deviation about a non-zero mean. In the case of the combined $p_{out}$ for all hadron-photon pairs within the specified $p_{T}$ ranges, for fragmentation photons the rms is found to be $0.787\pm0.022(stat)\pm0.031(sys)~\mathrm{GeV/c}$, while the inclusive is $0.562\pm0.0006~\mathrm{GeV/c}$, indicating a significantly broader distribution for fragmentation photons. The results shown for the inclusive and decay photons are consistent with previous measurments of $j_{T}$ for hadrons associated with a high-$p_{T}$ $\pi^{0}$ \cite{jets}, modulo effects of the decay kinematics resulting in a small decrease in the rms. This trend can also be seen in Figure \ref{pout_rms}, where the rms vs trigger $p_{T}$ is shown. These provide a direct measurement of the parton shower, a poorly understood process, without the added complication of the further fragmentation of the photon.

\end{document}